\documentclass{article}   
\usepackage{amscd,amsmath,amssymb,verbatim}
\usepackage{latexsym}
\usepackage{amsthm}
\usepackage{amsfonts}
\usepackage[left=2.5cm,top=3.0cm,right=2.5cm]{geometry} 
\usepackage[pdftex]{color,graphicx}
\usepackage{graphicx}
\usepackage{setspace}
\usepackage{multicol}
\usepackage[draft]{hyperref} 
\title{Segmented Vortex Telescope and its Tolerance to \\
Diffraction Effects and Primary Aberrations.\\
{\small {Accepted Jan. 24th 2013 scheduled for for publication in the OE Vol. 52, No. 8.}}}

\author{Juan P. Trevi{\~n}o,  Omar L\'opez-Cruz, and Sabino Ch\'avez-Cerda \thanks
{All authors: Instituto Nacional de Astrof\'isica, Optica y Electr\'onica, 
Apartado Postal 51/216, Puebla, M\'exico 72000. First author contact address: trevinojp@inaoep.mx}}

\begin{document}
\maketitle

%
\begin{abstract} 
\noindent
We propose the segmented Large Millimeter Telescope (LMT/GTM),as the largest spatial light modulator capable of producing vortex beams of integer topological charge. This observing mode could be applied for direct exoplanet searches in the millimeter or submillimeter regimes. We studied the stability of the vortex structure against aberrations and diffraction 
effects inherent to the size and segmented nature of the collector mirror. In the presence of low order aberrations the focal distribution of the system remains stable. Our results show that these effects depend on the topological charge of the vortex and the relative orientation of the aberration with respect to the antenna axis. Coma and defocus show no large effects in the image at the focal plane, nevertheless the system is very sensitive to astigmatism. Heat turbulence, simulated by random aberrations, shows that the system behaves in a similar way as astigmatism dissociating the vortices. We propose the Segmented Vortex Telescope as a novel approach for the detection of giant planets outside circumstellar disks around nearby stars. Since our results are applicable to other facilities with segmented surfaces, we suggest that this idea should be considered as a regular observation mode complementary to interferometric methods.
\end{abstract}

keywords: aberrations, astronomy, diffraction, millimeter waves, phase modulators, planets, spatial light modulators, telescopes.

\begin{multicols}{2}
\section{Introduction}
\label{sec:intro}
 
In general, waves with phase singularities and a rotational flow  around the singularity are called vortices. 
Vortex fields arise in nature \cite{Berry2000} and  enjoy a large number of technological applications. 
Vortex fields are found in physical systems covering the widest range of scales, from single photons to  astronomical black holes \cite{Tamburini2011}.  
Applications of vortex fields have been found in  optics, acoustics, telecommunications, astronomy, and other fields \cite{padg11}.

In astronomy, light vorticity has attracted a lot of attention since it has shown to be an elegant and efficient way  to reject light from a bright source on axis, allowing the  study of  its  surroundings. 
{Remarkable results have been obtained by the direct imaging of exoplanetary systems using ground-base optical telescopes equipped with coronagraphs~\cite{Swartzastro, pid, ovc, agphc, serabynnature,swartzelliptic}.}	
    
Vortex fields have been recently introduced to increase the number of independent channels available to transmit data signals.
By reshaping a common parabolic antenna, a phase vortex can be induced on a transmitted wave, creating a set of channels which are independent to those of a regular antenna~\cite{encTamb,radiovort}.  
The number of channels that can be transmitted using a single frequency is, in principle, unbounded. 
 This application has the potential of revolutionizing telecommunications. 

In this work, by combining the ideas above, we explore the feasibility of turning the Large Millimeter Telescope (LMT/GTM) \cite {gtm} into a coronagraph at millimeter wavelengths by reshaping its segmented collector mirror into a vortex generator. 
The main purpose is to investigate the potential of this system as a regular observational mode able to detect giant exoplanets or giant proto-exoplanets outside circumstellar disks. 
Given the  lower luminosity contrast  of such objects in millimeter wavelengths, this device would be very suitable for this application. 
We have investigated the stability of the system against diffraction effects induced by the segmented nature of the primary surface and possible damage to its segments. 
We took into account the effects of aberrations inherent to such large  structure, as those produced by deformations on the primary mirror due to its own weight or by temperature variations on its 50 meters diameter surface and its surroundings~\cite{Kaercher00}.  

\section{Optical Vortices and Phase Modifying Devices}
\label{sec:pmd}
A vortex coronagraph  is created by placing a phase modifying device at the focal plane of an optical system  \cite{Swartzastro,ovc, agphc,serabynnature,swartzelliptic}. 	%
At a relayed pupil plane most of the radiation from on axis sources is concentrated in a sharp ring that can be easily filtered away by means of a stop known as Lyot stop.
This setup is useful because it allows to filter out light from on axis sources. 
Radiation from off axis sources remains almost unchanged when it goes through the system. 
In this way it is possible to explore the vicinity of very bright objects  and even to produce images of objects dimmer and angularly close to the primary one.

For telescopes at millimeter wavelengths, any component intended to be implemented at the focal plane poses a technical challenge, since any element at this plane will attenuate the already low intensity signals. 
Additionally, detection  devices need to be kept at low temperatures to maximize their sensitivity \cite{bolocam}, so any additional element absorbs some of the incident radiation and  requires cooling which implies a higher cost of the entire system. 
Therefore, a vortex phase mask at the focal plane would not be the best option for radio and millimeter signals, hence we explore  the possibility to locate the vortex on the primary surface.

The LMT/GTM  has a Cassegrain design that operates at millimeter wavelengths~\cite{gtm}. 
Our proposal is to reshape  the segmented primary surface  to produce vortices at the working wavelengths of this telescope, which are 1.1, 1.4 and 2.1, mm.  \cite{AzTEC1, AzTEC2}.
The independent control of the segments in the primary mirror can be configured to produce integral order vortices at such wavelengths.  

Generating a vortex at the exit pupil creates a doughnut-like field distribution at the focal plane. 
Such setup is the reverse of  the signal transmission implemented in  \cite{encTamb, radiovort}. The detection device currently used at the LMT/GTM  is a bolometer array composed of 144 pixels~\cite{AzTEC1}. The beam size is reduced before it reaches the bolometer array. 
This way if the vortex distribution had a larger radius than the bolometer array, a further beam reduction would be necessary, however, this adjustment is straightforward.
This configuration was first explored as a means to reject light form bright sources by opening a window to observe incoherent background. 
In principle, the achievable contrast depends on the vortex order and the size of a field stop located at the focal plane. 
For two incoherent sources, the achievable contrast could be of the order of $~10^5$, so it is possible to detect Super-Jupiters as we will see below.
{Although not common in nature, for a pair of coherent sources, both on axis and off axis sources will come across the same vortex. However, some basic rules can be given for the detection of secondary sources \cite{pid}.}

\section{The Vortex collector mirror}
\label{sec:imaging}

An optical vortex is a wavefield with an azimuthally varying phase factor of the form $e^{im\varphi }$. The number $m\neq 0$ must take only integral values and is known as the topological charge of the vortex. 
The argument $m\varphi $ in the exponential represents an azimuthal linear variation of the phase {which} generates a helical wavefront. The corresponding Poynting vector field follows helical trajectories as well.
These kind of wavefields are known to carry Orbital Angular Momentum (OAM) \cite{padg11,Allen}. 
As a consequence, part of the radiation in the center of the beam is driven away from the propagation axis and a dark core is created at its center. 

A Cassegrain telescope like the LMT/GTM is described by a complex pupil function whose mathematical representation includes information about the optical power and aberrations of the system.
If the vortex is introduced at the exit pupil of the telescope, the mathematical expression of the field in this plane is 
\begin{equation}\label{eq:field}
U_{0}(r,\varphi )=T_{0}(r,\varphi )\,\exp\left\{-ik\left( \frac{r^{2}}{2f}\right)\right\}
\,e^{ikW(r,\varphi)}\,e^{im\varphi },
\end{equation}
where $W(r,\varphi)$ is the aberration function and $f$ is the focal distance of the mirror.
The wavenumber is given by $k=2\pi /\lambda $ for the wavelength $\lambda$,  while $m$ is an integer defining the topological charge of the vortex.

The comparative sizes between the working wavelengths and the gaps of the LMT/GTM are such that diffraction effects can be neglected. 
However, diffraction effects caused by gaps between segments have been investigated here including different gap sizes and configurations.

The field {$U$} at planes approaching the neighborhood of the focal plane $z=f^{-}$ is obtained by a Fresnel diffraction integral, while at the focal plane $z=f$ the integral transforms into the Frounhofer diffraction integral given by
\[
U_{f}(\rho ,\theta )={\mathcal{C}}\int_{-\pi }^{\pi } 
\int_{0}^{\infty}T_{0}(r,\varphi )\,
e^{ikW(r,\varphi)}\,e^{im\varphi} \]
\begin{equation}\label{eq:fft}
\exp\left\{i\frac{2\pi }{\lambda f}r\rho \cos
(\varphi -\theta )\right\}r\,dr\,d\varphi,
\end{equation}
where ${\mathcal{C}}$ is a complex constant related to the amplitude of the field. 
For a continuous mirror and in the total absence of aberrations, the field at the focal plane is radially symmetric and it reduces to the Hankel transform of order $m$ of the transmission function $T_{0}(r)$.  
In the general case that we will study here a 2D Fourier transfrom will be necessary as the circular symmetry of $T_0$ may be  broken by arbitrary aberrations or diffraction effects due to large gaps and damaged panels.
For non symmetric cases, we evaluate numerically { equation (\ref{eq:fft}) by means of a Fast Fourier Transform algorithm. 
In order to calibrate our numerical simulation we evaluated first the $m=0$ case and other cases for which analytical solutions are available, with known parameters. 
For other values of m we also checked the phase of the focal field for consistency.}

\section{Reshaping the LMT/GTM Primary Surface}
\label{vortex}

In the original design the collector mirror of the LMT/GTM has  a  50$m$ diameter and a radius {of curvature} of 35$m$. 
It is composed by 180 panels arranged in five rings. 
The innermost ring is formed of twelve segments, the second ring has 24, and the remaining three are composed by 48 {panels}. 
The length of each panel is 5$m$ and the corresponding angular extent depends on the ring where the segment is located. 
Currently, only the first three inner rings are operational, the corresponding diameter is of $32$m. 
It is feasible to implement our study to the current size of the primary,  hence we have  adjusted our simulations to the current diameter (see Fig.~1). 
Considering the full 50m aperture is straightforward.  

Currently, each segment of the primary mirror in the LMT/GTM is controlled by three actuators with a maximum displacement of $5,200\mu$m. 
The total wavefront error of the LMT/GTM design is of 75$\mu$m rms which means $\epsilon<\lambda /20$ at $1.5$mm.
The collector alone should have a 55$\mu$m rms error which represents $\epsilon \approx \lambda / 18$ at one millimeter. For longer wavelengths the error decreases.
The secondary mirror is a hyperbolic surface with a diameter of $2,570$mm  which represents a central obscuration of less than 10\% the radius of the 
exit pupil. 
The secondary has six degrees of freedom. 

To create a vortex of order $m$ at the collector mirror, the reflected converging wave must have an additional azimuthal phase $m\varphi $. 
For this reason, the collector must have a linear azimuthal deformation with a maximum of $m\lambda /2$ additional to its parabolic profile. 
For every ring, each one of the $N$ junctions between sectors should be displaced by $\frac{m\lambda }{2} \frac{n}{N}$, where $n=0,1,\dots N$, and $N$ is the number of segments in the corresponding ring. 
The maximum possible displacement of the segments allows for the creation of an $m=5$ vortex at $\lambda=2mm$. 
However, if the desired vortex order and corresponding wavelength required a maximum displacement larger than the maximum possible for a single segment, there is an option of creating a multiple pitch vortex, as shown in Fig~2. 
This configuration gives the same reflected wavefront but the physical displacements required are smaller.  

It is worth mentioning that these calculations are independent of the number of segments in the collector mirror because the surface produced has a ramp-like profile in the azimuthal direction, which in turn produces a true spiral wavefront.
Deforming the primary surface of the LMT/GTM as proposed here does not induce aberrations, it reshapes the wavefront into a convergent helicoid affected by diffraction due to segment gaps.
The rigidity of the collector mirror segments might cause errors when achieving the value of $m$, however the surface accuracy is below the error the system tolerates (approximately $5\%$) and they can be neglected.
This procedure to produce vortices might be compared to a step spiral phase mask \cite{maritamb} with $N$ steps. The latter approximates the ideal spiral wavefront by superposing $N$ plane waves with a relative retardation. This approximation causes a dependence on $N$ for such device, as opposed to the deformation of the primary segmented mirror studied here.

\section{Stability analysis to diffractive perturbations}
\label{sec:diffraction}
 As mentioned before, in the LMT/GTM the gaps are sufficiently small compared to  the wavelengths at which the system operates, nevertheless we present simulations to evaluate the diffraction effects that would appear for larger gaps. 
We split the problem to study  radial and angular gaps separately.

We first evaluate pupils formed from several continuous rings having a variable separation. 
The transmittance function of such a pupil is represented as a sum of circ functions.
Given the linearity of the Fourier transform, the focal field of this multiple ring pupil is a sum of the corresponding Fourier transforms. 
For any value of $m$, the effect on the focal distribution is very similar to what happens with a simple  annular pupil: the focal distribution shrinks radially, and the main maximum decreases relative to the secondary maximum as the radius of central obscuration increases (Fig.~3). 
This effect is interpreted as having an increase of resolution of the system  but a contrast decrease. 
For different values of $m$ the maximum field amplitude values drop almost linearly as the size gap increases. 
Nevertheless, the profile appears to be a scaled version of the original one, therefore the resulting distribution can be always approximated by a power law $r^m$  in the vicinity of the vortex core.

Next we will consider the case in which a complete sector of angle $\varphi$ is lost or darkened. 
For topological charges $m>1$ an obscured sector causes the vortex to dissociate and generate $m$ dark regions in the focal image.  
The topological charge of a vortex must be preserved along propagation, therefore these dark regions are associated to individual vortices with unitary charge. 
In general, vortices with opposite charge might cancel each other upon propagation while vortices with the same charge might travel alongside each other for long distances. Detailed information on vortex propagation might be found in~\cite{ellipticvortices, optvortprop}. 
In the case presented here vortex dissociation  happens as the $m$ order phase singularity separates into $m$ first order singularities as shown in Fig.~4.  A similar effect happens when the system is illuminated by two separate coherent sources~\cite{pid}. 
{However, the relative position of the dark regions due to the existence of multiple sources and that due to aberrations behave differently.}

Vortex dissociation is clearly appreciated if we observe the phase of the field, but it could be difficult to see with intensity detectors because the phase information is lost and vortices manifest as local minimma. However we might define a criterion to determine vortex dissociation depending on the contrast resolution of the detection device. 

The vortex dissociation effect is {also} observed for damaged segments as small as $\varphi\sim10^\circ$  which is a very small value for the angular extent of the real  LMT/GTM segments. 
In this case the vortex structure is unstable because symmetry is broken (Fig.~5).  
The focal distribution can be recovered to some extent if the opposite  segment is darkened since some symmetry is recovered. 
We can see in the figure that for $m=4$ the original profile is recovered in the vertical section. 
The case for $m=2$ is not fully recovered, however the horizontal section remains with the same profile but a lower contrast. 

Next, we examine when many opposite sectors are lost or obscured as shown in Fig.~6. 
A transverse section  of the {amplitude} at the focal plane shows that although the maximum decreases, the profile of the central dark area caused by the vortex remains unchanged. 
The same effects happen for the second or third rings composed of more segments. 
In these outer rings there is a larger  number of segments that contribute to the focal field, and although the obscured proportion might be high, the opposite reflecting sectors reconstruct a more symmetrical pattern.

If the missing panels are considered to be radial as shown in the figure, the maximum {field amplitude} decreases linearly for $m=0$, but for larger values of $m$,  the behavior departs slightly from linearity. 
In all cases the drop is proportional to the non-reflecting area. 

We have analyzed diffraction effect due to gaps size and found out that the central annular field is stable up to a ten to one ratio between the panel and the gap. 
In the LMT/GTM the real proportion is much smaller than the aforementioned ratio, for this reason, the effect can be neglected.

We now consider the effects of  damaged panels. We model this situation by leaving out randomly selecting panels (Fig.~7).  
We observe again that the basic vortex structure due to the azimuthal phase is maintained.  
Similar situations involving partially obstructed  optical vortices have shown that this type of fields are stable upon propagation \cite{forbes}. 
Even the absence of a complete ring, which is a special case of concentric ring pupil was considered. 
It can be seen that this can have small effects on the expected image (Fig.~7). 

\section{Stability analysis under aberrations}
\label{sec:aberrations}
So far we have analyzed how robust the segmented vortex telescope can be  considering diffraction perturbations due to gaps between panels or assuming damages of the collector surface.  
We now  draw the attention to the presence of aberrations e.g. Seidel  or balanced aberrations~\cite{mahajan1,mahajan2}.  
Typically, the atmospheric turbulence is modeled by a distribution of aberrations. 
For the LMT/GTM, in addition to atmospheric turbulence, 
gravitational effects cause deformations of the collector mirror and also
temperature fluctuations across its 50 meters diameter surface  cause aberrations~\cite{Kaercher00}. 
For this reason, we analyze the focal distribution of the vortex telescope
given Seidel and balanced aberrations  and also random phase aberrations.

We introduce the  wave aberration of the system as a factor $e^{ik W (r,\varphi )}$, with $W (r,\varphi)=A_j\Phi_j(r,\varphi )$ and $\Phi_j$ could be either a Seidel or a balanced (Zernike) aberration~\cite{mahajan1,mahajan2}. 
The aberration coefficient $A_n$ is given in units of the wavelength $\lambda$. 
We used Seidel spherical aberration, Seidel and Zernike astigmatism, and  finally Seidel and Zernike coma. 
Different aberration coefficients were selected to evaluate the stability of the central vortex structure at the focal plane as in the previous  cases. 
The aberration coefficients were chosen according to acceptable Strehl  ratio criteria \cite{mahajan3} selecting coefficients $A_i\leq 0.30$.
Symmetric aberrations such as defocus or first and second order spheres cause variations in the contrast of the focal image. 
Nevertheless the central profile of dark regions remains the same in the sense that  a power law fit in the radial variable is still possible in the vicinity of the vortex core. 
If the symmetry is broken by aberrations, the vortex dissociates in a similar way as with perturbations due to diffraction. 
Astigmatic aberration given by  $W=r^2\cos{\phi}$  generated visible vortex dissociation for low values, $A_a=0.03$. 
Nevertheless, the proportion of the inner maxima to the main  ring was about $5\%$. We set this proportion here to define a visible dissociated vortex.
Vortex dissociation occurs in an angle $\pi/4$ from the astigmatism axis, and a section perpendicular to this keeps the $r^m$ power law profile. 
The simulations include transversal sections of the focal distribution of amplitude to show how the zeros are separated depending on the value of $A_a$. 
For larger values of astigmatism the radiation distribution is deformed into a set of intensity lobes related to Hermite-Gaussian modes.
It is known that Hermite-Gaussian modes can be converted to and from  Laguerre-Gaussian (L-G) modes by means of an astigmatic mode converters. 
The latter are a type of vortex wavefields \cite{courtial}.
The system's response to aberration of coma $(W=r^3\cos{\phi})$ was more stable since no vortex dissociation was observed for aberration coefficients $A_c<0.35$ while the critical value for a $0.8$ Strehl ratio is $A_c=0.21$. 

A system with randomly damaged panels and a Seidel aberration combined  shows that the effects observed are closer to those produced by aberration alone. 
The dark region at the focal plane dissociates for aberration coefficients similar to those without segment loss. 
Fig.~9 shows the collector mirror with obscured panels and $A_c=0.15$ along with the focal image and transverse section. 
The image shows vortex dissociation and small diffraction effects.

For the large area of the collector mirror it is expected that inhomogeneities of air density caused by atmospheric fluctuations or heat turbulence, will generate a high order random aberration of the incoming wavefront.
{The thermal design of the LMT/GTM allows to evaluate and compensate deformations of the primary due to heat turbulence~\cite{Kaercher00}. 
However, the small fluctuations might not be possible to compensate.}
In order to investigate the influence of this type of perturbation on the system, we produced a surface with  random distributions of various frequencies (Fig. ~10).
Each case was modeled for Seidel aberrations and the aberration free case. 
In each case we found out vortex dissociation for a phase aberration of a quarter wavelength measured peak to peak (Fig.~11).
The sensitivity of the system to this type of aberration requires a careful calibration and wavefront compensation of the system. 

Since the antenna follows objects in the sky, astigmatism caused by deformation of the collector mirror due to gravity, is constantly changing. If we wanted to distinguish between the presence of aberrations and double coherent sources, we might have to analyze the behavior of the dark regions {as the antenna scans a given object in the sky}. 
Aberrations caused by changes in the antenna position would cause the dark regions to move as the antenna moves. On the other hand, if the dark regions remain unchanged as the scanning goes on, it would mean that there is a double source. {More precise schemes are being investigated by our group at the moment, and a report will be developed.}

\section{Astronomical Application}
\label{sec:astronomical}

Consider the LMT/GTM working at 1.1 mm, at this wavelength the field of view  is $1.5^\prime$ (minutes of arc) and a beam size of  $5^{\prime\prime}$ (seconds of arc) with a pointing accuracy is  $1^{\prime\prime}$. The specified final surface accuracy of 70$\mu$m, this results in an Antenna Efficiency of 46\% at this wavelength.  
{In the mm regime the contrast between the luminosity of a solar-like host star ($T_{eff}\sim6000$ K) and a Super-Jupiter ($T_{S-J}\sim1700$ K, $R_{S-Jupiter}\sim 1.5 R_{Jupiter}$, see e.g.~\cite{Carson})  becomes of the order of $10^3$, while in the optical regime the contrast could be several times higher, 
$\sim10^8$. 
However, cold  dust in circumstellar disks becomes the dominant source of radiation at 1 mm, hence the contrast between a face-on circumstellar disk (mean $T_{Disk}\sim 50$K, see below for average size) e.g., \cite{Hughes} and
a Super-Jupiter could be as large as $10^{-6}$. These are rough figures under the assumption that the host star, the circumstellar disk, and the planet emit as black bodies. 
 
The mode of the diameter distribution  of  circumstellar disks~\cite{catalogue} is about 300 AU (1 Astronomical Unit= 149,597,870.691 km). 
A  disks of about this size,  would appear unresolved at   60 pc (1 parsec= 206265 AU); hence we can use LMT/GTM in the SVT mode as a coronagraph to search for exoplanets  orbiting around star-circumstellar disks systems
at distances larger than 60 pc. The LMT/GTM's field of view would  cover  separations larger than 5000 
AU from the host star.  Guided by previous observations~\cite{Wright}, we found that  exoplanetary systems have shown a wide variety of configurations. 
Massive planets have been found to orbit their host stars at wide orbits, ranging from several hundreds to over a thousand AU, from their host stars~\cite{Ireland}. 
A comprehensive survey for massive planets at wide orbits could help to impose constraints on the origin of exoplanets and their relation with the environment~\cite{Perets}.
Super-Jupiter have properties similar to low mass brown dwarf, hence LTM/GTM working as SVT could also help to study the frequency of brown dwarfs in multistellar systems. 

The geometry of circumstellar disks can be explored using sub-Rayleigh resolution~\cite{tambOvercoming} in classically resolved, or unresolved circumstellar disks, or in the case of stellar systems with multiple components (stars and planets), this mode can increase the resolution up to  $0.5^\prime$ at 1.1mm, which can only be attained with interferometric arrays. 
Moreover, the high mapping speeds (0.55 ${\rm deg}^2/{\rm hr}/{\rm mJy}^2$ for LMT/GTM with AzTEC) of large single dish telescopes make them more efficient than interferometric arrays. Therefore, we suggest that future large millimeter single dish  telescope should consider SVT as an regular  observational mode, which would guide follow up observations with interferometric arrays. }

\section{Conclusions}
\label{sec:conclusions}

We have investigated the properties of the Large Millimeter Telescope as the largest spatial electromagnetic wavefront light modulator capable of producing vortex beams of integer topological charge at selected wavelengths.
We studied its stability against  diffraction and aberration perturbations.

Diffraction caused by gaps between segments and by possible mechanical damages of the LMT/GTM panels
is not critical since the main features of the vortex structure at the focal plane remain unaffected by the segmented nature of the telescope even in highly perturbed cases. 

Regarding to aberrations, for small values of defocus and coma, the focal distribution of the system remains stable.  As expected, we observed that the system is very sensitive to astigmatism. Non-radially symmetric aberrations cause the vortex to dissociate into $m$ vortices, and as a result, dark zones appear in the focal distribution. 
The values at which this happens depend on $m$ and on the relative orientation of the aberrations. 
The dark zones of a single source always appear perpendicular to the aberration axis. 

For aberrations with random wavefront distribution, simulating  heat turbulence, the system shows a high sensitivity which must be taken into account since even single sources produce vortex dissociation generating similar focal distributions as those produced by two coherent sources.
In general, single sources produce multiple dark regions at the focal plane when the symmetry of the system is broken either by diffraction  or by the  presence of non-symmetric aberrations. 

The LMT/GTM working as a SVT could offer unprecedented  opportunities~${}$for exoplanetary studies.  The SVT can help  to search for giant  planets (Super-Jupiters)  outside debris disks \cite{wolf2012} and also work in the sub-Rayleigh \cite{tambOvercoming} mode to resolve multiple components.  

\section*{Aknowledgements}

The authors would like to acknowledge the reviewers of this manuscripts. Their valuable comments helped us to improve significantly our work.\\

The first author would like to acknowledge Alfredo Agustín Monta\~na Barbano for the valuable sources for the development of this work. He would also like to acknowledge Jesus Emmanuel Gomez-Correa for fruitful discussion.


\begin{thebibliography}{10}
\bibitem{Berry2000}
M.~V. Berry.
\newblock Making waves in physics.
\newblock {\em Nature}, 403:21, 2000.
\bibitem{Tamburini2011}
F.~Tamburini, B.~Thid\'e, G.~Molina-Terriza, and G.~Anzolin.
\newblock Twisting light around black holes.
\newblock {\em Nature Physics}, 7:195.
\bibitem{padg11}
A.~M. Yao and M.~J. Padgett.
\newblock Optical angular momentum: origins, behavior and applications.
\newblock {\em Advances in Optics and Photonics}, 3:161.
\bibitem{Swartzastro}
Jr. G.~A.~Swartzlander, E.~L. Ford, R.~S. Abdul-Malik, L.~M. Close, M.~A.
  Peters, D.~M. Palacios, and D.~W. Wilson.
\newblock Astronomical demonstration of anoptical vortex coronagraph.
\newblock {\em Opt. Express}, 16(14):10200--10207, Jul 2008.
\bibitem{pid}
Jr. G.~A.~Swartzlander.
\newblock Peering into darkness with a vortex spatial filter.
\newblock {\em Opt. Lett.}, 26(8):497--499, Apr 2001.
\bibitem{ovc}
G.~Foo, D.~M. Palacios, and Jr. G.~A.~Swartzlander.
\newblock Optical vortex coronagraph.
\newblock {\em Opt. Lett.}, 30(24):3308--3310, Dec 2005.
\bibitem{agphc}
D.~Mawet, P.~Riaud, O.~Absil, and J.~Surdej.
\newblock Annular grove phase mask coronagraph.
\newblock {\em Astrophysical Journal}, 633(2).
\bibitem{serabynnature}
E.~Serabyn, D.~Mawet, and R.~Burruss.
\newblock An image of an exoplanet separated by two diffraction beamwidths from
  a star.
\newblock {\em Nature: Letters}, 464:1018--1020.
%
\bibitem{swartzelliptic}
G. J. Ruane and G. A. Swartzlander
\newblock Optical vortex coronagraphy with an elliptical aperture.
\newblock {\em Appl. Opt.}, 2:52, 2013
\bibitem{encTamb}
F.~Tamburini, E.~Mari, A.~Sponselli, B.~Thid\'e, A.~Bianchini, and F.~Romanato.
\newblock Encoding many channels on the same frequency through radio vorticity:
  First experimental test.
\newblock {\em New Journal of Physics}, 14:033001, 2012.
\bibitem{radiovort}
B.~Thide, F.~Tamburini, E.~Mari, F.~Romanato, and D.~Barbien.
\newblock Radio beam vorticity and optical angular momentum.
\newblock {\em arXiv}, (arXiv:1101.6015v1 [astro-ph.IM]), 2011.
\bibitem{gtm}
W.M. Irvine, E.~Carrasco, and I.~Aretxaga.
\newblock {\em The Large Millimeter Telescope: Neighbors Explore the Cosmos}.
\newblock 2005.
\bibitem{Kaercher00}
Hans~J. Kaercher and Jacob W.~M. Baars.
\newblock Design of the large millimeter telescope/gran telescopio millimetrico
  (lmt/gtm).
\newblock pages 155--168, 2000.
\bibitem{bolocam}
G.~Jason et.al.
\newblock Bolocam: a millimeter-wave bolometric camera.
\newblock {\em Proc. SPIE, Advanced Technology MMW, Radio, and Terahertz
  Telescopes, Thomas G. Phillips; Ed.}, 3357:326.
\bibitem{AzTEC1}
K.~S. Scott, J.~E. Austermann, T.~A. Perera, G.~W. Wilson, I.~Aretxaga, J.~J.
  Block, D.H. Huges, S.~Kim Y.~Kang, P.~D. Mauskopf, D.~B. Sanders,
  N.~Scoville, and M.~S. Yun.
\newblock The aztec mm wavelength camera.
\newblock {\em Mon. Not. R. Astron. Soc.}, 386.
\bibitem{AzTEC2}
G.~W. Wilson, J.~E. Austermann, T.~A. Perera, K.~S. Scott, P.~A.~R. Ade, J.~J.
  Block, J.~Glenn, S.~R. Golwala, Y.~Kang S.~Kim, D.~Lydon, P.~D. Mauskopf,
  C.~R. Predmore, C.~M. Roberts, and M.~S. Yun.
\newblock Aztec millimetre survey of the cosmos field - I. data reduction and  source catalogue.
\newblock{\em Mon. Not. R. Astron. Soc.}, 385, 2225-2238 (2008)
\bibitem{Allen}
L.~Allen, M.~W. Beijersbergen, R.~J.~C. Spreeuw, and J.~P. Woerdman.
\newblock Orbital angular momentum of light and the transformation of
  laguerre-gaussian laser modes.
\newblock {\em Phys. Rev. A}, 45:8185--8189, Jun 1992.
\bibitem{maritamb}
E.~Mari, G.~Anzolin, F.~Tamburini, M.~Prasciolu, G.~Umbriaco, Antonio.
  Bianchini, C.~Barbieri, and F.~Romanato.
\newblock Fabrication and testing of l $=$ 2 optical vortex phase masks for
  coronography.
\newblock {\em Opt. Express}, 18(3):2339--2344, Feb 2010.
\bibitem{ellipticvortices}
S.~Ch\'{a}vez-Cerda, J.~C. Guti\'{e}rrez-Vega, and G.~H.~C. New.
\newblock Elliptic vortices of electromagnetic wave fields.
\newblock {\em Opt. Lett.}, 26(22):1803--1805, Nov 2001.
\bibitem{optvortprop}
Guy Indebetow.
\newblock Optical vortices and their propagation
\newblock {\em Jour. Mod. Opt.}, 40:1, 1993. 
\bibitem{forbes}
R.~Rop, I.~A.~Litvin, and A.~Forbes.
\newblock Generation and propagation dynamics of obstructed and unobstructed
  rotating orbital angular momentum-carrying helicon beams.
\newblock {\em J. Opt.}, 14(3):035702, 2012.
\bibitem{mahajan1}
V.~N. Mahajan.
\newblock Zernike circle polynomials and optical aberrations of systems with
  circular pupils.
\newblock {\em Appl. Opt.}, 33(34):8121--8121, Dec 1994.
\bibitem{mahajan2}
V.~N. Mahajan.
\newblock Zernike annular polynomials and optical aberrations of systems with
  annular pupils.
\newblock {\em Appl. Opt.}, 33(34):8125--8127, Dec 1994.
\bibitem{mahajan3}
V.~N. Mahajan.
\newblock {\em Optical Imaging and Aberrations Part I}.
\newblock SPIE Optical Engineering Press, 1998.
\bibitem{courtial}
J.~Courtial and M.J. Padgett.
\newblock Performance of a cylindrical lens mode converter for producing
  laguerre gaussian laser modes.
\newblock {\em Optics Communications}, 159(1 3):13 -- 18, 1999.
\bibitem{Carson}
J. Carson, et. al.
\newblock Direct Imaging Discovery of a `Super-Jupiter' Around the late
B-Type Star $\kappa$ And
\newblock{\em arXiv}, (arXiv:1203.6271v1 [astro-ph.IM]).
\bibitem{Hughes}
A. M. Hughes, D. J. Wilner, 
C. Qi, C and M. R. Hogerheijde.
\newblock Gas and Dust Emission at the Outter Edge of Protoplanetary Disks.
\newblock  {\em Astrophysical Journal}, 678(1119), 2008.
\bibitem{catalogue}
Karl Stapelfeldt
\newblock Catalogue of Circumstellar Disks. 
\newblock in {\it Catalogue of Circumstellar Disks}.
\newblock Retrieved January 2013.
\newblock http://www.circumstellardisks.org/index.php
\bibitem{Wright}
J. T. Wright, O. Fakhouri, G. W. Marcy, E. Han, Y. Feng, J. A. Johnson, A. W. Howard, D. A. Fischer, J. A. Valenti, J. Anderson, and N. Piskunov.
\newblock{The Exoplanet Orbit Database}
\newblock{\em arXiv}, (arXiv:1012.5676v3 [astro-ph.SR]
\bibitem{Ireland}
M. J. Ireland, A. L. Kraus, F. Martinache, N. M. Law, and L. A. Hillenbrand
\newblock Two wide planetary-mass companions to solar-type stars in upper scorpius
\newblock {\em Astrophysical Journal}, 762(113).
\bibitem{Perets}
H. B. Perets and M. B. N. Kouwenhoven.
\newblock On the origin of planets at very wide orbits from the recapture of free floating planets
\newblock {\em Astrophysical Journal}, 750(83).
\bibitem{tambOvercoming}
F.~Tamburini, G.~Anzolin, G.~Umbriaco, A.~Bianchini, and C.~Barbieri.
\newblock Overcoming the rayleigh criterion limit with optical vortices.
\newblock {\em Phys. Rev. Lett.}, 97:163903, Oct 2006.
\bibitem{wolf2012}
Circumstellar disks and planets.
\newblock {\em arXiv}, (arXiv:1211.3744v1 [astro-ph.SR]).

\end{thebibliography}

\end{multicols}


\clearpage
\section*{Authors}
\begin{itemize}
\item Juan Pablo Trevino:
Is a PhD candidate at Instituto Nacional de Astrof\'{\i}sica, Optica y Electr\'onica. His main research interests are the mathematical modeling of light modal propagation and its applications to imaging systems as the human eye and telescopes. He is also interested in the modeling and compensation of aberrations with adaptive optics systems for various applications. 

\item Omar Lopez-Cruz:
Omar López-Cruz is a researcher at Instituto Nacional de Astrof\'{\i}sica, Optica y Electr\'onica, and visiting scientist at the Astrophyisics Group of the University of Bristol in the UK.  
His research interests range from stellar astrophysics, extragalactic astronomy, and observational cosmology.  For this aim, he   uses all sort of telescopes, interferometric arrays,  and 
virtual  observatories.  He is leading a Mexican team of scientists and engineers to build 
a single-antenna experiment to search for the signs of the formation of the first stars 
in the universe using  the 21cm hyperfine transition of neutral hydrogen as a tracer. 

\item Sabino Ch\'avez-Cerda:
Is a full time researcher at Instituto Nacional de Astrof\'{\i}sica, Optica y Electr\'onica in the Optics department. He obtained his PhD degree from the Imperial College and has been recently appointed as  Fellow of the Optics Society of America. His main research fields are propagation of non diffractive fields and rigorous mathematical physics aspects of optics.
\end{itemize}

\clearpage

\section*{Figures}

\begin{figure}[h]
	\centering
		\includegraphics[width=12cm]{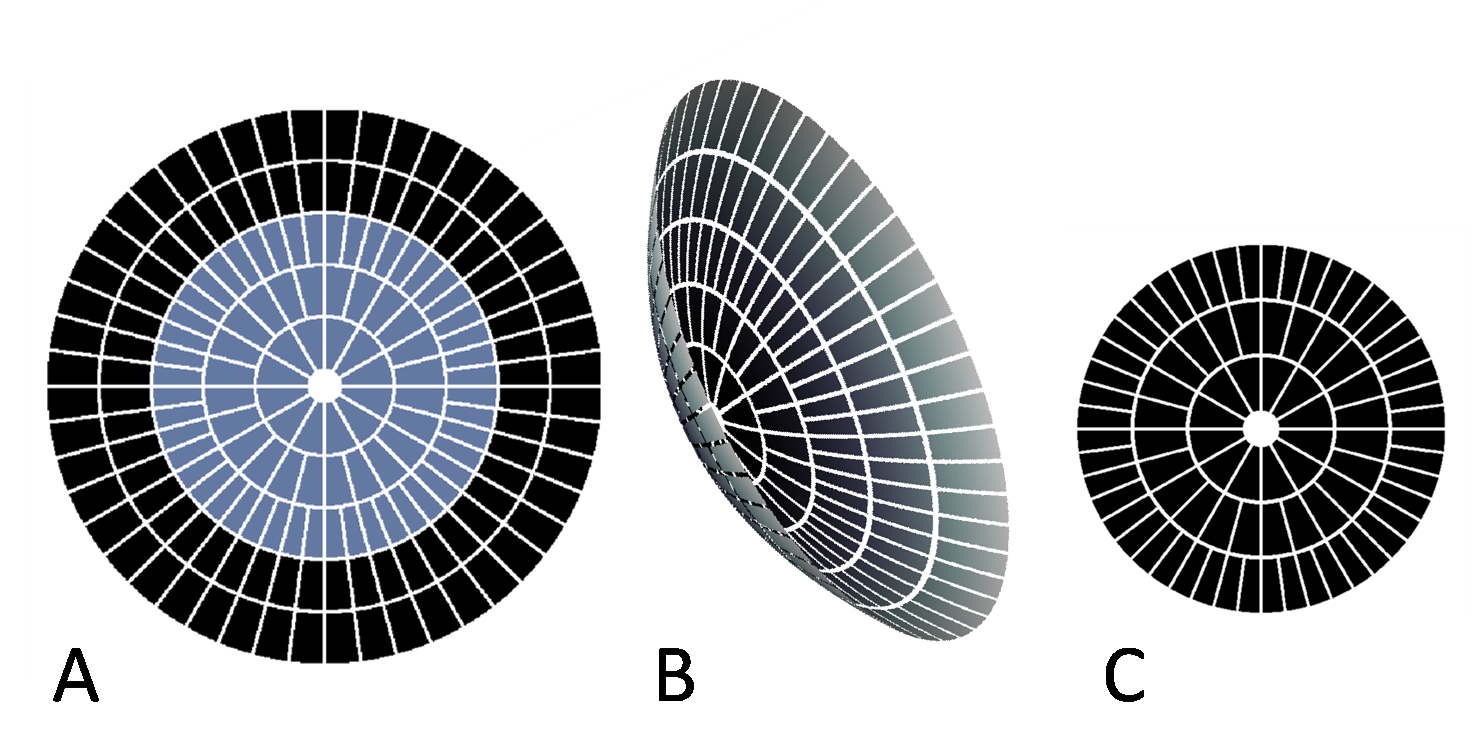} 
	\caption{ A: Complete setup with five rings as originally designed.   Segments in black are currently not active. B: The collector mirror of the LMT/GTM generates a waveftont with this shape. The lines correspond to the gaps between the telescope segments. C: Front view of the transmittance function of the current three rings of the collector. }
\end{figure}
\begin{figure}[h]
	\centering
		\includegraphics[width=10cm]{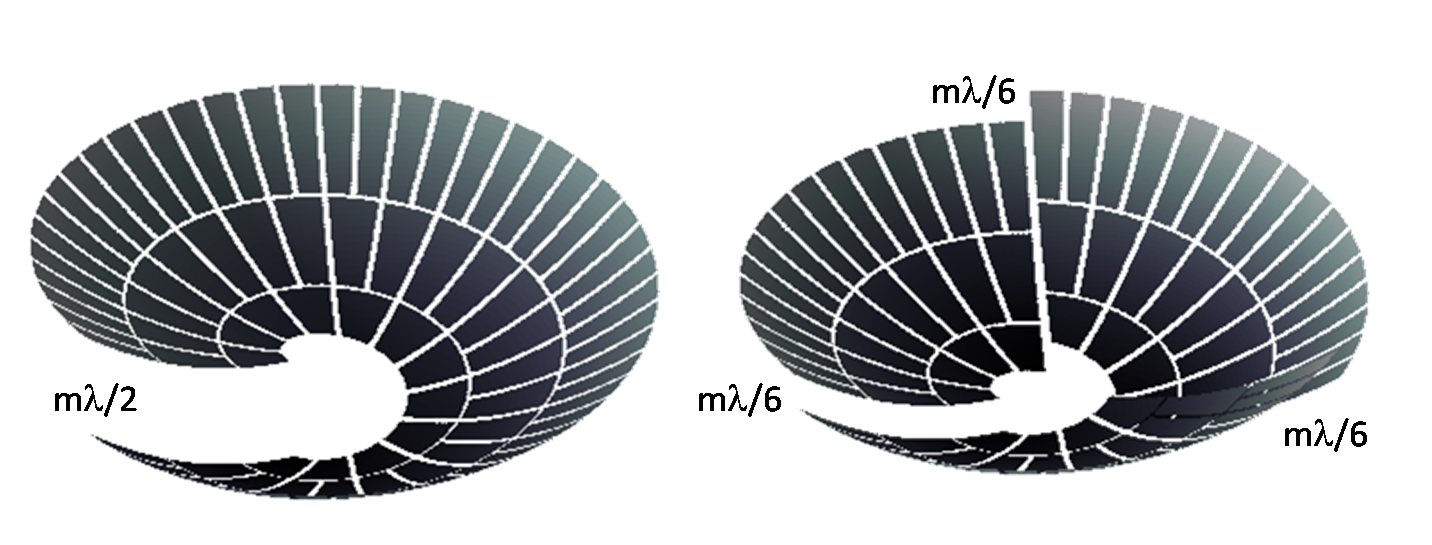} 
	\caption{ Multiple pitch mirror for the segmented vortex telescope. }
\end{figure}
\begin{figure}[h]
	\centering
		\includegraphics[width=12cm]{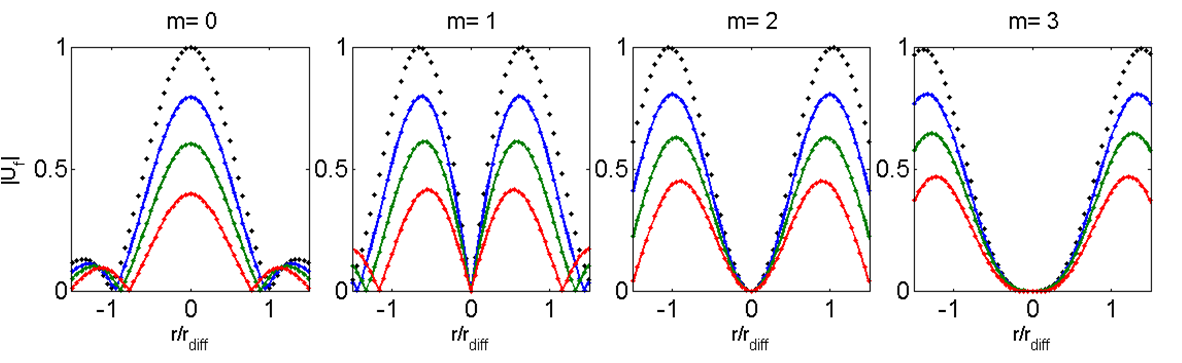} 
	\caption{  Transverse section of the normalized field amplitude at the focal plane  for a pupil with three rings. The different lines show how the intensity drops as the gap between rings increases. The maximum amplitude as a function of the gap size ratio decreases almost linearly. }
 \end{figure}
\begin{figure}[h]
	\centering
		\includegraphics[width=8cm]{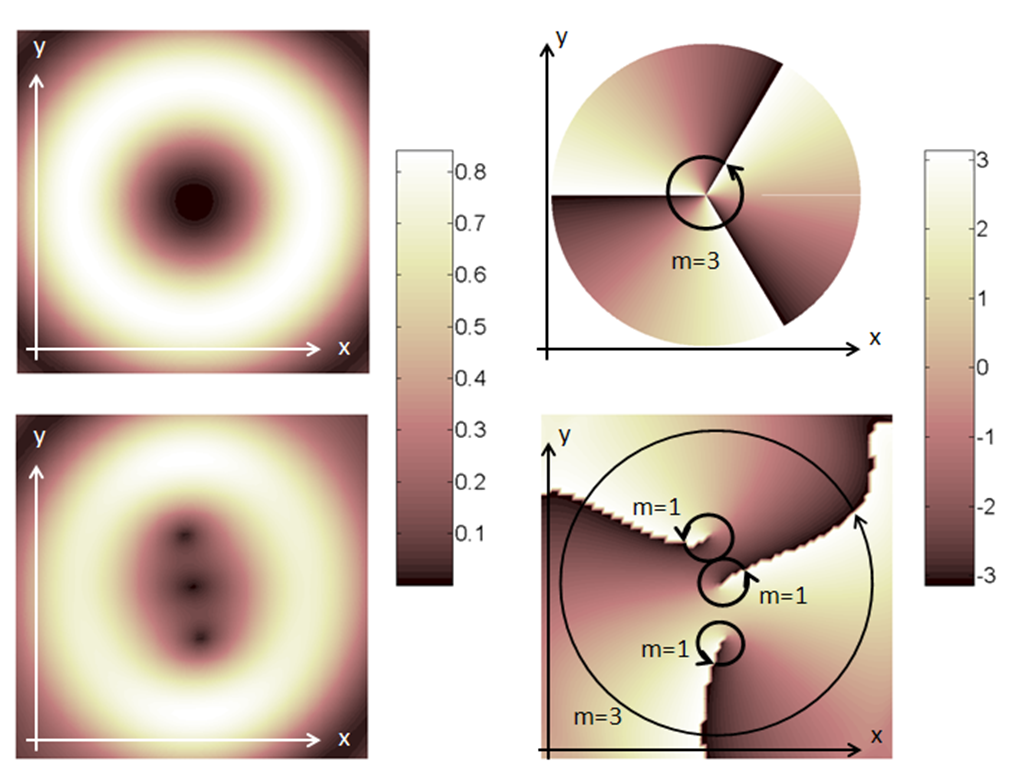} 
	\caption{  Figure shows phase and normalized amplitude at the focal plane for an $m=3$ vortex (top) dissociated into three $m=1$ vortices (bottom).  }
\end{figure} 
\begin{figure}[h]
	\centering
		\includegraphics[width=12cm]{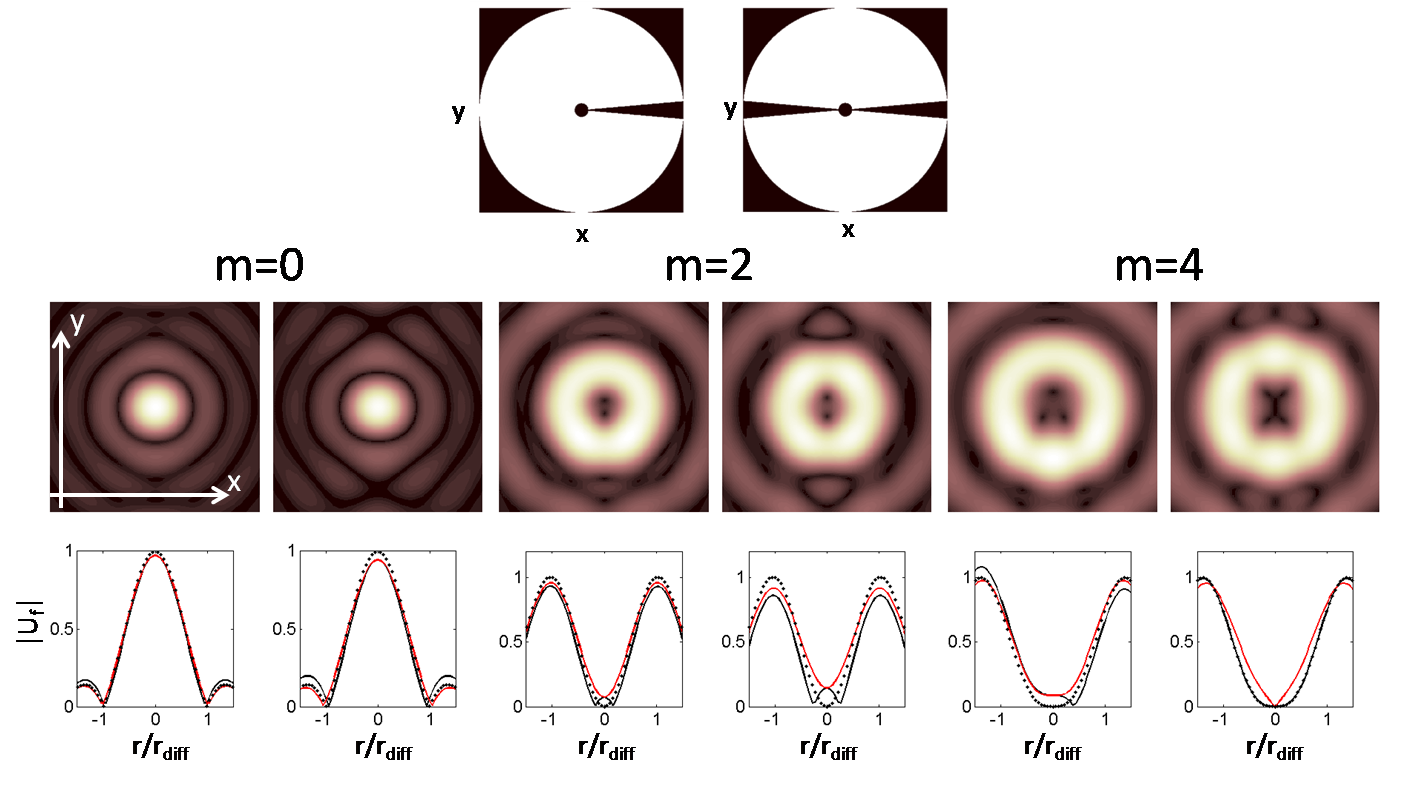} 
	\caption{ A complete damaged sector breaks the symmetry of the system and causes vortex dissociation for small damaged angles. The top images represent the obscured sectors. Second and third row show the normalized field amplitude distribution and a cross section respectively for $m=0$, $m=2$, and $m=4$ (bottom) imaged with a damaged sector of $\phi=15^\circ$.  
The dotted line is the normalized amplitude with a non damaged pupil, while the solid lines are vertical (black) and horizontal (red) cross sections of the amplitude with damaged pupils. Notice the dips that correspond to the dissociated vortices.}
 \end{figure}
\begin{figure}[h]
	\centering
		\includegraphics[width=12cm]{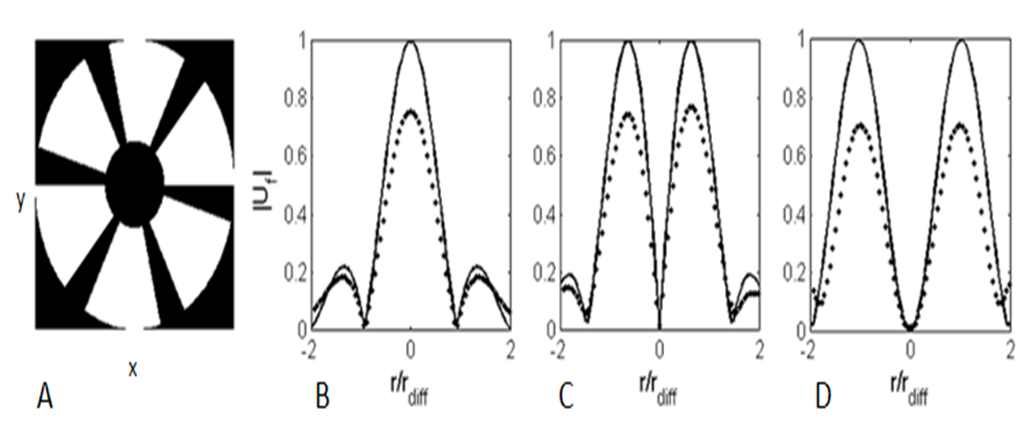} 
	\caption{ A. Segmented ring pupil with gap ratio  of $1/4$. B. $m=0$ 
C. $m=1$, and D. $m=3$. The solid line is the normalized  field amplitude for a continuous ring, while the dotted line is for the segmented one.
The intensity field drops as the gap size ratio increases, but the shape of the field is preserved.  }
\end{figure}
\begin{figure}[h]
	\centering
		\includegraphics[width=12cm]{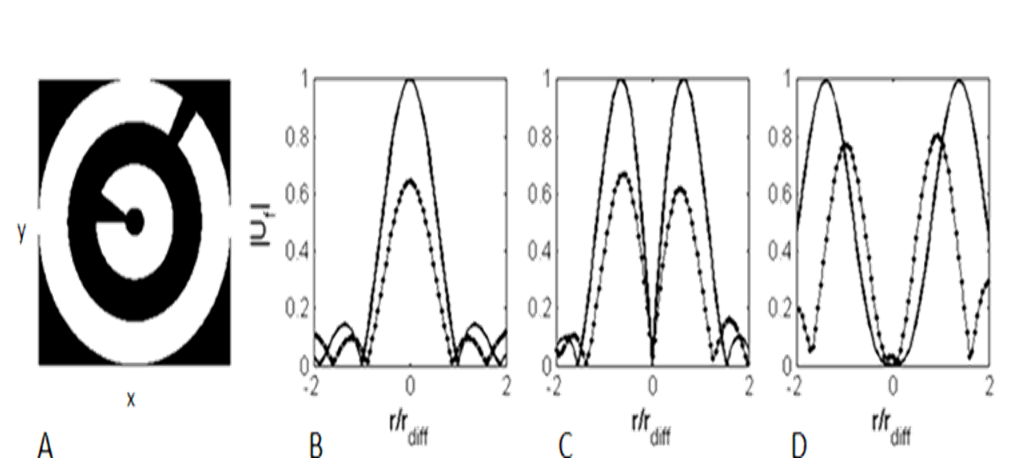} 
	\caption{ A. A pupil with full ring and random segments damaged. B. $m=0$ C. $m=1$, and D. $m=3$. The solid line is the field intensity for a  continuous ring, while the dotted line is for the damaged one.
The intensity field drops  but the shape of the field is preserved.   }
\end{figure}
\begin{figure}[h]
	\centering
		\includegraphics[width=8cm]{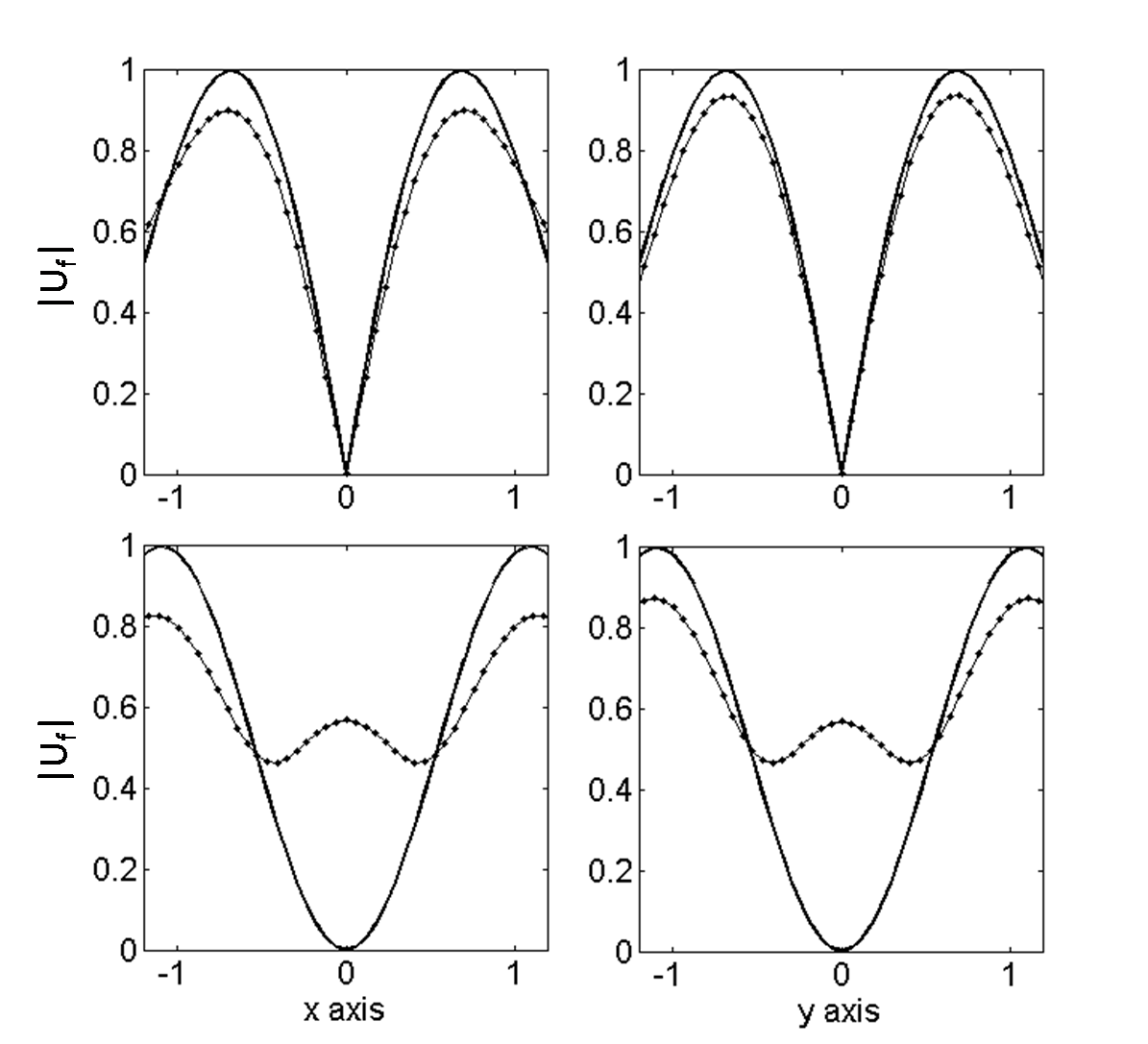} 
	\caption{ Cross sections of intensity at the focal plane for pupil with astigmatic aberration $A_a=0.3$ (dashed lines) and the unaberrated reference (solid line). Top: $m=1$, Bottom: $m=2$. }
 \end{figure}
\begin{figure}[h]
	\centering
		\includegraphics[width=12cm]{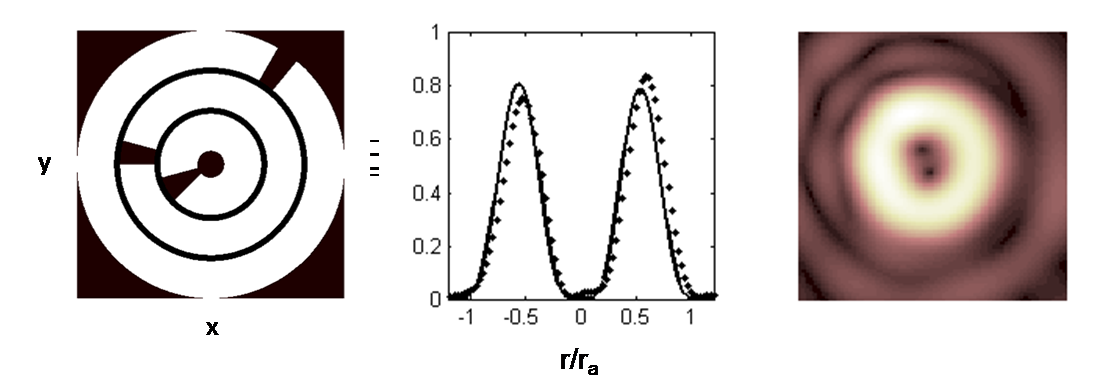} 
	\caption{  A case combining damaged random segments and aberration of coma. The solid line represents the ideal case. }
\end{figure}
\begin{figure}[h]
	\centering
		\includegraphics[width=8cm]{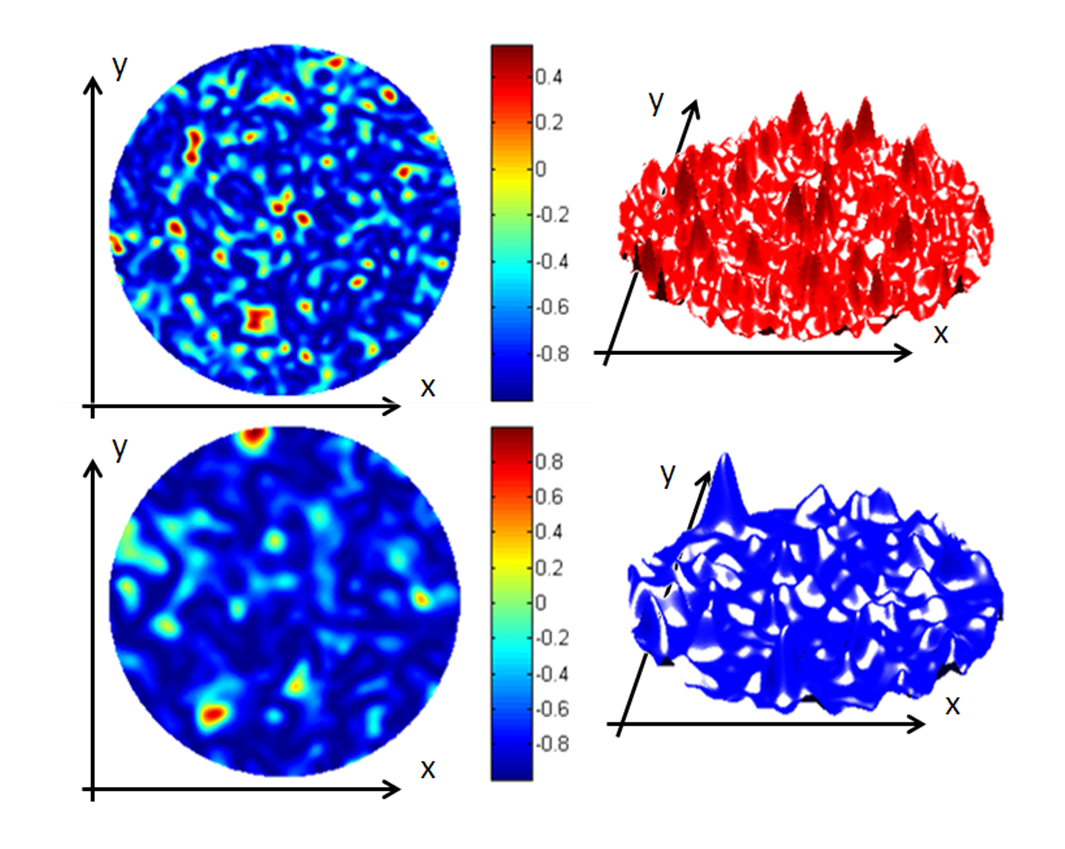} 
	\caption{  Wavefront aberration surface generated from a random frequency distribution. Top: High frequencies. Bottom: Low frequencies. }
 \end{figure}
\begin{figure}[h]
	\centering
		\includegraphics[width=13cm]{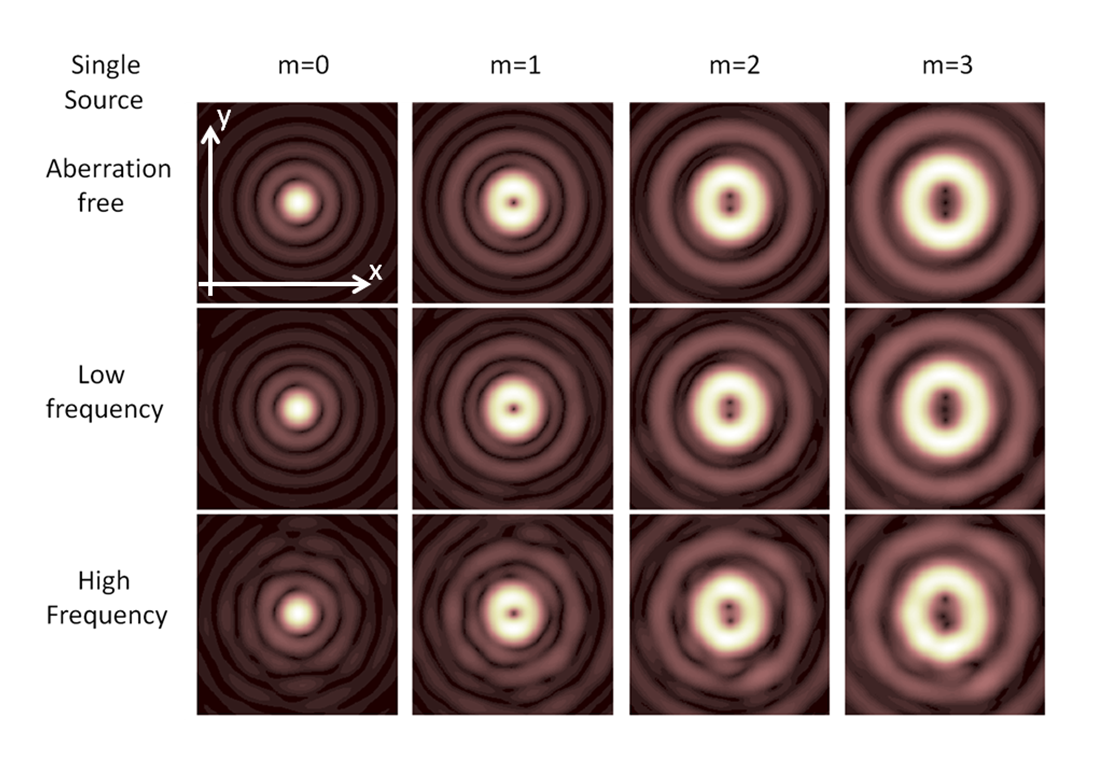} 
	\caption{ Fluctuations generating a random phase aberration of a quarter wavelength peak to peak dissociates vortices.}
\end{figure}

\end{document}